\documentclass{nature}

\usepackage[USenglish]{babel}

\usepackage{amssymb}   
\usepackage{amsmath}
\usepackage[T1]{fontenc} 
\usepackage[latin1]{inputenc}
\usepackage{lmodern} 
\usepackage{color}
\newcommand{\ket}[1]{\left|#1\right\rangle}

\usepackage{graphicx,xcolor}
\usepackage{amsmath, bbm}
\usepackage{hyperref}
\usepackage{epstopdf}
\usepackage{bbm}
\usepackage{soul}
\newcommand{\mean}[1]{\langle #1\rangle}

\renewcommand{\imath}[0]{\mathrm{i}}

\newcommand{\tensor}[1]{\overline{\overline #1}}					

\title{Engineering long-lived vibrational states for an organic molecule}
\author{Burak Gurlek$^{1,2}$, Vahid Sandoghdar$^{1,2}$ \& Diego Martin-Cano$^{1,3}$ }

\date{\today}

\begin{document}

\maketitle

\begin{affiliations}
 \item Max Planck Institute for the Science of Light, 91058 Erlangen, Germany 
 \item Department of Physics, Friedrich Alexander University, 91058 Erlangen, Germany.
 \item Departamento de Física Teórica de la Materia Condensada and Condensed Matter Physics Center (IFIMAC), Universidad Autónoma de Madrid, E28049 Madrid, Spain;
\end{affiliations}

\begin{abstract}
The optomechanical character of molecules was discovered by Raman about one century ago \cite{Raman-1928}. Today, molecules are promising contenders for high-performance quantum optomechanical platforms \cite{Roelli-2016,Benz-2016, Neuman-2019, Roelli-2020} because their small size and large energy-level separations make them intrinsically robust against thermal agitations. Moreover, the precision and throughput of chemical synthesis can ensure a viable route to quantum technological applications. The challenge, however, is that the coupling of molecular vibrations to environmental phonons limits their coherence to picosecond time scales  \cite{Basche-2008}. Here, we improve the optomechanical quality of a molecule by several orders of magnitude through phononic engineering of its surrounding. By dressing a molecule with long-lived high-frequency phonon modes of its nanoscopic environment, we achieve storage and retrieval of photons at millisecond time scales and allow for the emergence of single-photon strong coupling in optomechanics. Our strategy can be extended to the realization of molecular optomechanical networks.
\end{abstract}

Molecules are usually considered in the realm of chemistry and as building blocks of organic matter. However, scientists have been increasingly turning their attention to molecules for their naturally rich and compact quantum mechanical settings, where a wide range of electronic, mechanical and magnetic degrees of freedom could be efficiently accessed and manipulated~\cite{Engel-2007, Roelli-2016, Wang-2019, Lin-2020, Bayliss-2020}. A particularly intriguing promise of molecules is their use as quantum optomechanical platforms \cite{Fiore-2011, Ramos-2013, Aspelmeyer-2014, Roelli-2016, Benz-2016, Neuman-2019, Roelli-2020, Reitz-2020}, but this idea confronts the challenge that the various molecular degrees of freedom quickly lose their ``quantumness'' when coupled to the phononic bath of the environment in the condensed phase. In this theoretical work, we show how to create long-lived phononic states by tailoring the vibrational modes of organic crystals that embed impurity guest molecules. 

\begin{figure}
	\begin{center}
		\includegraphics[width=13.5cm]{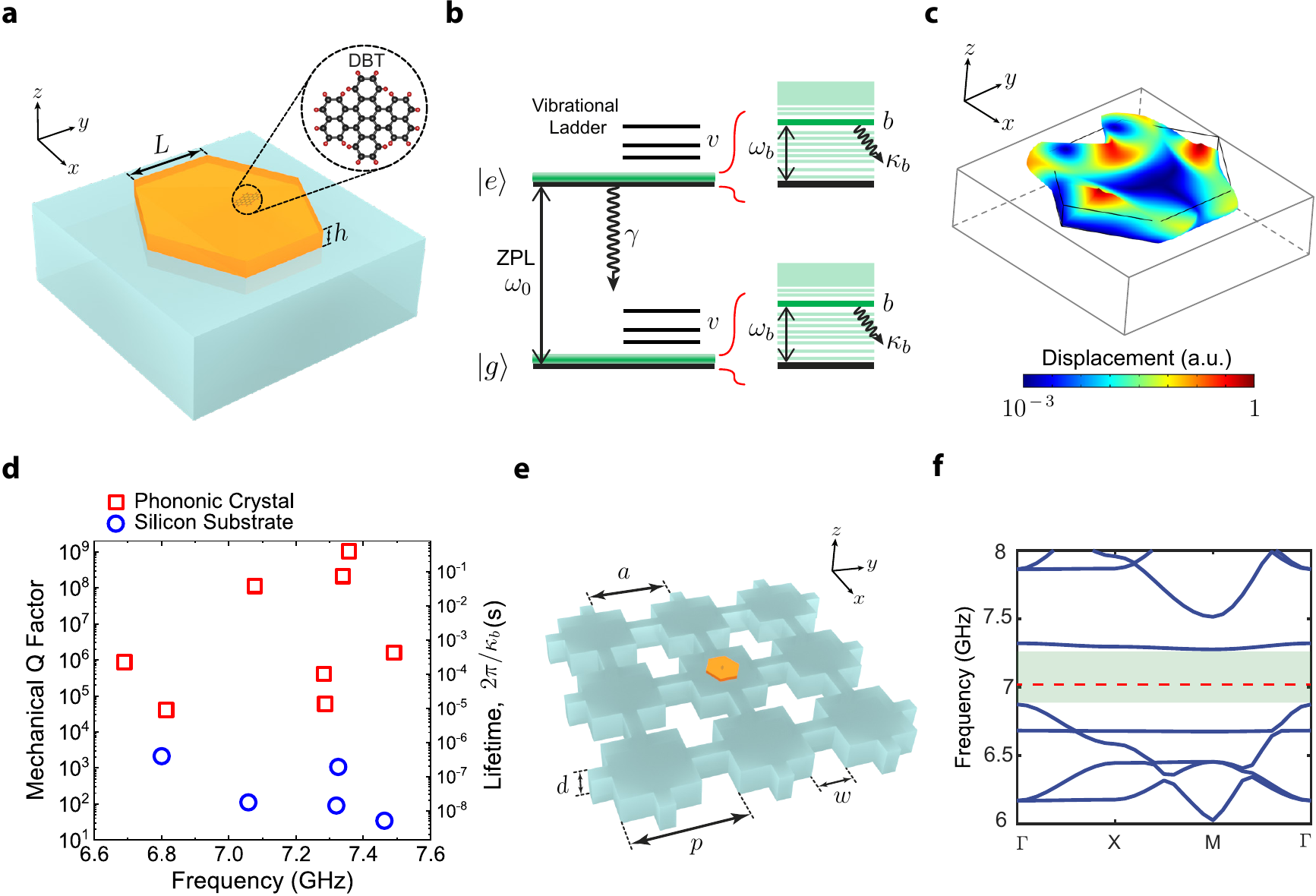}
	\end{center}
	\caption{\textbf{a,} Illustration of an organic molecule (DBT) embedded in a nanocrystal placed on a macroscopic substrate. For simplicity, the crystal is assumed to be a hexagonal prism with height $h$ and edge length $L$. \textbf{b,} Simplified Jablonski diagram of a single dye molecule with electronic ground ($\ket{g}$) and excited ($\ket{e}$) states, including coupling to its own vibrons ($\ket{v}$), lattice phonons in continuum (light green lines) and a long-lived phonon mode ($\ket{b}$, green lines). \textbf{c,} Displacement field profile of a nanocrystal mode on a silicon substrate at resonance frequency $\omega_b/2\pi=$7.02~GHz. \textbf{d,} Quality factor (left vertical axis) and lifetimes (right vertical axis) of the nanocrystal phonon modes when placed on an unstructured silicon substrate (blue) and on the silicon phononic crystal (red) shown in (e). \textbf{e,} The nanocrystal is placed on a silicon phononic crystal made of cross-shaped holes with lattice constant $p=1.2\,\mu$m, cross width $w=360$~nm, cross length $a=1.166\,\mu$m, and membrane thickness $d=300$~nm. \textbf{f,} Corresponding simulated acoustic band structure along the high symmetry points inside the Brillouin zone. The full bandgap is highlighted in green. The dashed line displays the frequency of the nanocrystal mode shown in (c). 
	}
	\label{Schematic}
\end{figure}

Figure\,\ref{Schematic}(a) portraits a prototypical system based on organic polycyclic aromatic hydrocarbons (PAHs), such as an anthracene (AC) crystal as a host, containing dibenzoterrylene (DBT) dopant molecules \cite{Wang-2019}. Such solid-state matrices are manufacturable down to the nanometer scale by means of different methods~\cite{Gmeiner-2016,Pazzagli-2018,Hail-2019}, and their guest molecules are known to support excellent quantum coherent optical transitions~\cite{Basche-2008, Wang-2019}. By nature, a dye molecule establishes a rich optomechanical system with large cross sections for transitions involving its electronic states ($\ket{g, e}$) and vibrational levels (vibrons, $\ket{v}$)~\cite{Reitz-2020} (see Fig.\,\ref{Schematic}(b)). Vibrons can be long-lived in the gaseous state, but if the molecule is embedded in a solid matrix, the molecular levels also couple to the phonons ($\ket{b}$) in the host \cite{Reitz-2020,Clear-2020}. The large phononic density of states in macroscopic solids then result in fast decays of vibrons in the range of picoseconds~\cite{Basche-2008}.

Our strategy is to design the phononic landscape of the AC crystal and its substrate to create long-lived phonon modes for transferring and storing information from the guest molecule via external laser fields. To achieve this goal, we first reduce the crystal dimensions to gain access to discrete acoustic modes in the frequency range of a few GHz. This facilitates the selection of single modes in their ground state ($n_b\approx 0$ for $T\sim 0.1$\,K) and enhances the molecular electron-phonon coupling due to smaller phonon mode volumes (see estimations below). In a second step, we engineer the environment of the crystal to minimize the damping of its phonons. As a concrete working example, we place a hexagonal AC nanocrystal with height $h=50$\,nm and side length $L=200$\,nm on a silicon phononic crystal (PC) with a bandgap in the GHz frequency range~\cite{Maccabe-2020}.

Figure\,\ref{Schematic}(c) shows the results of numerical simulations (see Methods section (MS)) for the displacement field of a typical GHz-range vibrational mode of the AC nanocrystal on a macroscopic unstructured silicon substrate. The blue symbols in Fig.\,\ref{Schematic}(d) display the quality factors $Q$ and lifetimes of different phonon eigenmodes of the nanocrystal. Aside from some variations arising from different energy distributions, we find the mode lifetimes to be on time scales of nanoseconds~\cite{Meth-1989}, which happens to be similar to the electronic excited-state lifetime for this particular set of parameters, i.e., $\kappa_b\sim\gamma$. To inhibit the decay of the nanocrystal modes, we decrease the density of states in the underlying substrate by nanostructuring it into a PC~\cite{Maccabe-2020} with a bandgap centered about 7\,GHz (see Figs.\,\ref{Schematic}(e) and \ref{Schematic}(f)). The red symbols in Fig.\,\ref{Schematic}(d) report a dramatic increase of the phonon lifetimes to \textit{millisecond} timescales ($Q\sim 10^8$) accompanied by a slight modification of the AC nanocrystal mode frequencies. 

The dynamics of the interaction between a molecule and a single phonon mode of its environment can be described by the Hamiltonian~\cite{May-2004,Reitz-2020}
\begin{align}
\label{eqn:Hamiltonian}
H =\hbar \omega_{0}~\sigma^\dagger \sigma + \hbar\omega_b~ b^\dagger b + \hbar g_0~ \sigma^\dagger \sigma ( b^\dagger +b) .
\end{align}
Here, $\sigma$ and $b$ denote the electronic and vibrational annihilation operators, respectively, and the operators $\sigma^\dagger$ and $b^\dagger$ are their creation counterparts. The first two terms in Eq.\,(\ref{eqn:Hamiltonian}) correspond to the individual energies of a single electron and a phonon mode with transition frequencies $\omega_0$ and $\omega_b$, respectively. The last term provides the electron-phonon interaction, characterized by the strength $g_0$, which arises from linear displacements produced by the molecule in the crystal and related to the Debye-Waller factor $e^{-g_0^2/\omega_b^2}$ of a phonon mode~\cite{Reitz-2020}. Interestingly, this molecular interaction  mirrors the form of common cavity optomechanical  Hamiltonians~\cite{Aspelmeyer-2014,Ramos-2013,Puller-2013,Tian-2014,Roelli-2016,Neuman-2019} with the optomechanical constant $g_0=D\tilde{s}\sqrt{\omega_b/2\hbar EV} $ ~\cite{Ramos-2013}, where $D$ denotes the deformation potential induced in the impurity by the local strain $\tilde{s}$, $E$ is the Young modulus of the host, and $V$ is the phonon mode volume. Considering the deformation potential estimated in recent DBT:AC experiments~\cite{Clear-2020} ($D/2\pi\hbar\sim 1300~$~THz), the small volumes of the nanocrystals ($V\sim 2.5\times 10^{-4}~\mu\text{m}^3$), $\tilde{s}\approx 0.04-0.12$, and $E\approx 10^{10}~\text{Pa}$~\cite{Tian-2014} (see SI), one arrives at $g_0/2\pi \sim 50-150$~MHz, comparable with or larger than the electronic decay rates of typical quantum emitters~\cite{Basche-2008,Yeo-2014,Bhaskar-2020}. This provides access to laser-driven coherent optomechanical interactions and, particularly, coherent transfer of photons from a guest molecule to long-lived crystal phonons, which we explore below via master equation simulations (see MS).

\begin{figure}
	\begin{center}
		\includegraphics[width=14cm]{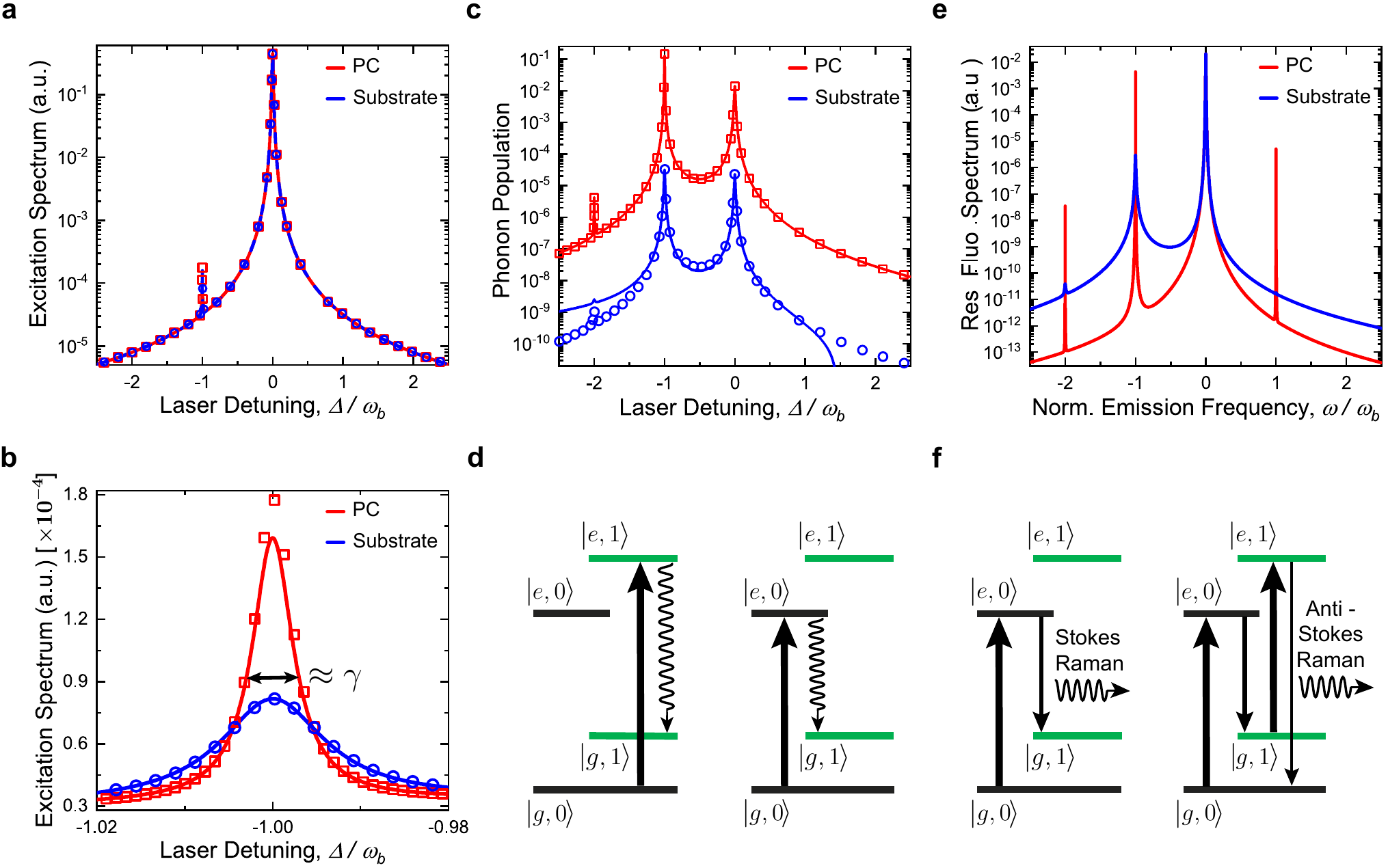}
	\end{center}
	\caption{Spectroscopy of a single molecule coupled to a macroscopic substrate (blue, $\kappa_b=1.6\gamma$) and a phononic crystal (red, $\kappa_b=10^{-3}\gamma$). Parameters: $\gamma/2\pi=40$~MHz, $g_0=\gamma$, $\omega_b=177.15\gamma$, $\Omega=\gamma$. \textbf{a,} Excitation spectrum ($\langle\sigma^\dagger \sigma\rangle$) versus laser frequency detuning ($\Delta=\omega_0-\omega_L$). \textbf{b,} A close-up of the peak at the vibrational transition in (a). \textbf{c,} Vibrational excitation spectrum ($\langle b^\dagger b\rangle$) versus laser frequency detuning. In (a-c), symbols represents the numerical results, and solid lines show the outcome of analytical formulas from adiabatic approximations between electrons and phonons (see more details in SI). \textbf{d,} Diagrams for the breakage of Kasha's rule in (c), where the molecule is excited via a vibrational level (left) and the zero-phonon line (right) \textbf{e,} Resonance fluorescence spectrum versus emission frequency ($\omega$). \textbf{f,} Diagrams of the different Stokes (left) and Anti-Stokes (right) scattering processes involved in (e).}   
	\label{excitation-spectrum}
\end{figure}

First, let us examine the electronic excited state population of the molecule via fluorescence ($\propto \langle\sigma^\dagger \sigma\rangle$) driven by a laser field at the onset of saturation, i.e. when Rabi frequency $\Omega$ is comparable to $\gamma$ (see MS). Figure\,\ref{excitation-spectrum}(a) displays this quantity for the AC nanocrystal on a silicon substrate (blue circles) and coupled to a PC that leads to $\kappa_b=10^{-3}\gamma$ (red squares) versus the laser frequency detuning. The dominant feature represents excitation via the zero-phonon line (ZPL), which connects $\ket{e, v=0}$ and $\ket{g, v=0}$ and is barely changed in the two cases due to the small Debye-Waller factor for a single vibration ($(g/\omega_b)^2\approx 10^{-5}$). A close look at the spectrum around the vibration frequency $\omega_b$ (see Fig.\,\ref{excitation-spectrum}(b)), however, shows that this transition is three times more strongly excited on the PC and its linewidth is limited by $\gamma$. This observation is also confirmed by analytical calculations performed under adiabatic approximations of the electronic and phononic fields, displayed by solid curves in Fig.\,\ref{excitation-spectrum}(a-c) (see MS). 

In Fig.\,\ref{excitation-spectrum}(c), we plot the population of phonons against the laser frequency detuning. We find that the generated signal is dramatically enhanced by 3 orders of magnitude in the presence of the PC. Moreover, comparing the heights of the two main resonances, we observe that the achieved phonon population is much higher if the excitation takes place through state $\ket{e, b}$, i.e., via the nanocrystal vibrational mode (see Fig.\,\ref{excitation-spectrum}(d)). The linewidth of this transition remains limited by $\gamma$ since its Einstein A-coefficient involves a similar frequency and dipole moment as for the decay of $\ket{e, b=0}$. These effects reflect the breakage of Kasha's rule, which states that fluorescence emission usually takes place from the lowest excited state~\cite{Kasha-1950}. The observed phenomena manifest that the molecule is dressed with the vibrational modes of its nanoscopic environment, acting on a par with the intrinsic molecular vibrational levels. We remark that stronger Rabi drivings or lower phonon decay rates would result in the appearance of optomechanical self-sustained oscillations \cite{Ludwig-2008}, the analysis of which goes beyond the scope of this paper. The onset of this regime is, however, highly suppressed for pulsed driving due to the finite duration of the excitation, as we exploit below for modes with longer phonon lifetimes.

A further illuminating way to investigate the fingerprints of long-lived phonons is to analyze the resonance fluorescence spectrum of the system driven at the ZPL ($\Delta=0$)~\cite{Wrigge-2008}. The resulting emission spectrum in Fig.\,\ref{excitation-spectrum}(e) shows very narrow peaks dominated by the phonon decay rate and a larger number of overtones. Remarkably, the spectrum also displays an anti-Stokes line at higher frequencies when coupled to the PC. We attribute this feature to coherent Raman scattering assisted by the vibrational levels of the electronic excited state in two steps. First, the vibrational manifold in the electronic ground state is coherently excited via Stokes processes through the ZPL. Second, its coupling to the upper manifold occurs via coherent anti-Stokes scattering (see Fig.\,\ref{excitation-spectrum}(f)), under moderate Rabi frequencies for long-lived vibrations ($\kappa_b\ll\gamma$). This phenomenon is the molecular analogue of the single-photon optomechanical strong-coupling regime, which yet remains to be reported experimentally in the solid state~\cite{Aspelmeyer-2014}. In a nutshell, the molecule acts as an optical nanoantenna \cite{Sandoghdar-2013} that facilitates its coherent optomechanical coupling to phonons when $g_0$ is large enough to exceed the mechanical and optical losses, much similar to the role of a cavity in conventional quantum optomechanics \cite{Aspelmeyer-2014}.

\begin{figure}
\begin{center}
\includegraphics[width=8cm]{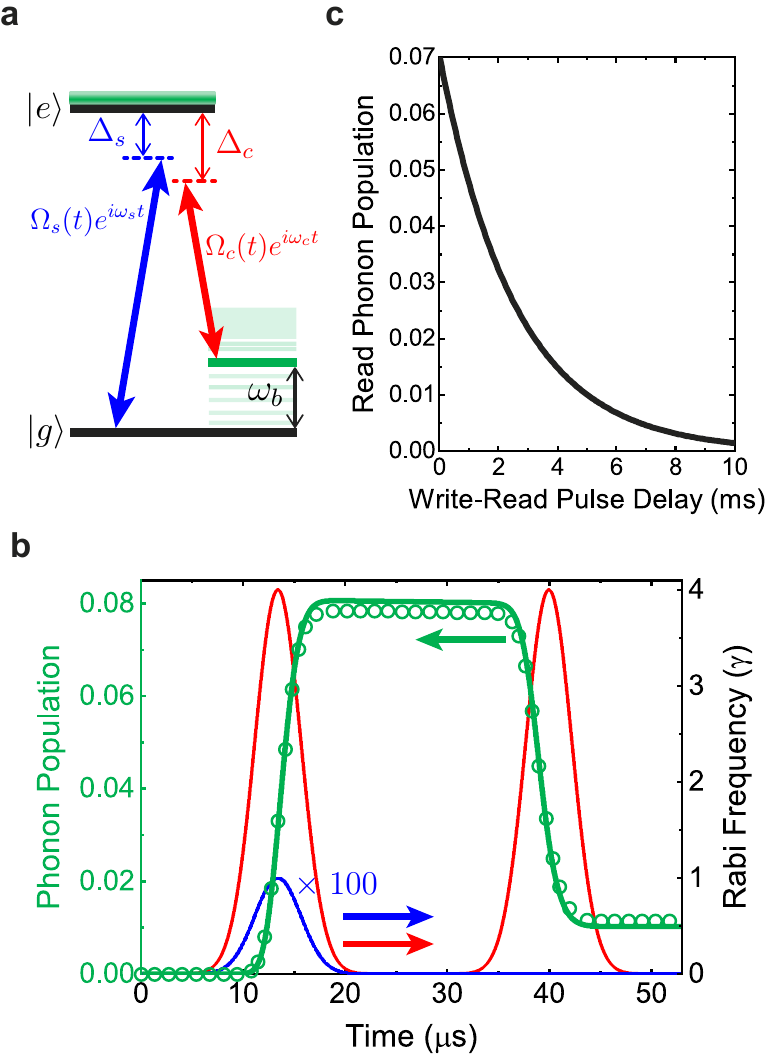}
\end{center}
 \caption{\textbf{a,} Energy level diagram for a single-molecule quantum memory based on a stimulated Raman scheme involving signal and control pulses with Rabi frequencies $\omega_{s(c)}$, amplitudes $\Omega_{s(c)}$ and frequency detunings $\Delta_{\text{s(c)}}=\omega_0-\omega_{s(c)}$ with respect to the ZPL. \textbf{b,}  The green symbols display the numerical results for the phonon population in the long-lived state ($\kappa_b=1.6\times 10^{-6}\gamma$) resulting from the pulse sequences shown in blue and red. The green solid line represents the analytical results based on a coherent model. The rest of the molecular parameters remains as in Fig.~\ref{excitation-spectrum}. The right vertical axis represents the synchronized signal (blue) and control (red) pulse sequences in units of the time-dependent Rabi frequency, given by $\Omega_{s(c)}(t)=\Omega_{s(c)}~\text{Exp}\left[-\frac{(t-t_0)^2}{\tau_p^2}\right]e^{j\omega_{\text{s(c)}}t}$ at $\Delta_{s,c}=0$ and pulse lengths of $\tau_p=5.3~\mu$s. \textbf{c,} The analytically calculated read phonon population versus delay between write and read pulse sequence.}
\label{Memory}
\end{figure}
      
Next, we exploit the long coherence time of the proposed molecular platform to realize a quantum memory. Here, we employ a strong control pulse to coherently map (write) a weak signal to the long-lived phonon mode by means of stimulated Raman scattering (see Fig.~\ref{Memory}(a)). The pulsed signal stored in the form of vibrations can then be coherently retrieved after a certain delay by applying a strong read pulse, as displayed in Fig.~\ref{Memory}(b). The green symbols in this figure show an example of the numerical results for the generation and read-out of the population for a phonon state with ms lifetime ($\kappa_b=1.6\times 10^{-6}\gamma$). We note that the excitation of phonons in Fig.~\ref{Memory}(b) is the result of a cooperative optomechanical driving of the molecule by both the signal and control pulses. Indeed, a single-beam excitation of phonons, either via control pulses without signal or vice versa, leads to two orders of magnitude smaller phonon population. We also estimated the efficiencies of the write and read steps to be $\eta_{\rm w}=40\%$ and $\eta_{\rm r}=86\%$, respectively. To do this, we first normalized the power of the transmitted control laser beam to the incident power (average of $0.04$ photons per pulse width of about $5\,\mu$s) and then integrated and compared the corresponding quantities obtained with and without the signal pulse (see SI for details). 

To obtain a deeper insight into the dynamics of our molecular memory system, we also developed an analytical model to solve the equations for coherent vibrational and electronic mean fields (see SI for details). The excellent agreement between the numerical and analytical calculations (solid curve) in Fig.~\ref{Memory}(b) confirms the coherent nature of the optical storage and read processes. The outcome of the analytical coherent model also yields compact efficiency expressions, $\eta_{\rm w}\approx\sqrt{\frac{8}{\pi}}\frac{1}{M_\text{w}}[\frac{1}{2}-e^{-M_\text{w}\sqrt{\frac{\pi}{2}}}+\frac{1}{2}e^{-M_\text{w}\sqrt{2\pi}}]$ and $\eta_{\rm r}\approx 1-e^{-M_\text{r}\sqrt{\frac{\pi}{2}}}$ (see SI for derivations), characterized by a universal memory constant $M_{\text{w}(\text{r})}|_{\tilde\Delta=0}=\frac{2g_0^2|\Omega_{c,\text{w}(\text{r})}|^2\tau_{p}}{\gamma|\gamma/2+i\omega_b|^2}$. The resulting maximum write and read efficiencies for this model amount to $40\%$ and $100\%$, respectively, close to our numerically found observations and show similar performance for a wide set of parameter, including smaller $g_0$ (see sweep maps in SI). A further benefit of the analytical model is that it allows us to directly and efficiently examine the full coherence time of the memory up to milliseconds, as presented in Fig.~\ref{Memory}(c).

In conclusion, we have shown that by sculpting the nanoscopic environment of a molecule, one can dress it with new vibrational modes and, thus, engineer a novel optomechanical quantum memory with coherence times in the order of milliseconds. Furthermore, we demonstrated that the efficient coupling of the electronic and vibrational degrees of freedom of the composite system leads to a regime in which the conventional Kasha's rule no longer holds. These phenomena allow one to enter single-photon strong coupling in optomechanics~\cite{Aspelmeyer-2014}, a paradigm that has not yet been observed in the condensed phase. Our strategy can be readily generalized to the design of hybrid quantum optomechanical platforms in which the large optical cross section of a quantum emitter is combined with tailored vibrational modes of its environment to access long-lived ground states, providing an attractive alternative to systems currently explored for quantum information processing based on spins~\cite{Bhaskar-2020, Pompili-2021}.

\newpage
\begin{methods}
\section*{Finite element simulations}

To gain insight into the acoustic phonon properties of AC crystals placed on unstructured and phononic crystal silicon substrates \cite{SafaviNaeini-2010}, we study the linear elastodynamic wave equation 
\begin{equation}
	\omega_n^2\rho(\mathbf{r})\mathbf{u}_n(\mathbf{r})+\nabla\cdot[\tensor{C}(\mathbf{r}):\tensor{s}_n(\mathbf{r})]=0\label{eq:eph_waveq}.
\end{equation}
Here, $\omega_n$ is the resonance frequency of the normal mode with displacement field vector $\mathbf{u}_n(\mathbf{r})$,  $\tensor{s}_n(\mathbf{r})=\nabla\mathbf{u_n}(\mathbf{r})$ represents its strain field tensor  and  $\tensor{a}:\tensor{b}$ denotes double dot product between dyadics $\tensor{a}$ and $\tensor{b}$. The studied system is characterized by the mass density $\rho(\mathbf{r})$ and  the spatial elastic stiffness tensor $\tensor{C}(\mathbf{r})$. Since AC possesses an anisotropic stiffness tensor~\cite{Dye-1989}, we use numerical simulations based on the finite-element method (COMSOL Multiphysics~\cite{COMSOL-2016}). To acquire access to the lifetime properties of the resulting modes, we apply absorbing perfectly matched layer (PML) and solve the resulting eigenvalue problem of Eq.~\ref{eq:eph_waveq} for the different materials (see more simulation details of the modes in Supplementary Information (SI)).

\section*{Open-systems dynamics}

In order to account for the properties of an open system~\cite{Gardiner-2004} in the molecular Hamiltonian of Eq.\,\ref{eqn:Hamiltonian}, we further include the Lindblad superoperators $\mathcal{L}\{\mathcal{O}\}{\rho}=\mathcal{O}\rho\mathcal{O}^\dagger-1/2\{\mathcal{O}\mathcal{O}^\dagger,\rho\}$ acting on the density matrix $\rho$, which account for the decay of the operators in the bath for the electronic transition $\mathcal{L}\{\sqrt{\gamma}~\sigma\}$ and the vibrational AC crystal mode $\mathcal{L}\{\sqrt{\kappa_b}~b\}$, with fullwidth decay rates $\gamma$ and $\kappa_b$, respectively. To pump the molecular Hamiltonian with one or two lasers, we include the driving terms $ \hbar\Omega(\sigma^\dagger e^{-i \omega_L t}+ \sigma e^{i \omega_L t})$ with Rabi frequency $\Omega$ and driving frequency $\omega_L$. For the coherent pulse studies, we apply this term with time-varying contributions from the signal and control Rabi frequencies $\Omega_{s(c)}$, with laser frequencies $\omega_{s}$ and $\omega_{c}$, respectively. We then solve the resulting master equation numerically via the QuTiP~\cite{Johansson-2012} to explore the main dynamical and steady-state properties of the system. To gain further insight into the simulations for the parameters considered in this work, we have also developed two analytical models based on the Quantum Langevin approach~\cite{Reitz-2019,Reitz-2020}. For the steady state averages, we noticed a fair approximation based on the decorrelated evolution between the electronic population and phononic displacement operator $\mean{\sigma^\dagger \sigma D(t)D^\dagger(t')}\approx \mean{\sigma^\dagger \sigma(t)}\mean{D(t)D^\dagger(t')}$, which together with the dominance of the terms $\mean{\sigma b},\mean{\sigma^\dagger b^\dagger}$ enable us to extend previous formulas~\cite{Reitz-2019,Reitz-2020} to driving conditions near saturation. For weak signals in the two-pulse driving for the memory study, we observe a significant coherent generation of electronic and phononic amplitudes ($\mean{O^\dagger O} \approx \left|\mean{O}\right|^2$). Such approximations allow us to solve the resulting equations with good accuracy and provide quite compact analytical expressions for the main driving scenarios here considered (see more derivation details in SI).

\end{methods}

 \addendumlabel{Authors' contributions: B.G. performed the calculations. D.M-C. and V.S. supervised the project. All authors contributed to writing the paper.}

\addendumlabel{Acknowledgements: This work was supported by the Max Planck Society. D.M-C. also acknowledges support from the fellowship (LCF/BQ/PI20/11760018) from ''la Caixa'' Foundation (ID 100010434) and from the European Union´s Horizon 2020 research and innovation programme under the Marie Sklodowska-Curie grant agreement No 847648. We thank Claudiu Genes for fruitful discussions. }

\bibliographystyle{naturemag}
\bibliography{Gurlek-02042021}

\end{document}